\documentclass[aps,prl,twocolumn,noshowpacs,superscriptaddress]{revtex4-2}

\usepackage{graphicx}% Include ure files
\usepackage{dcolumn}% Align table columns on decimal point
\usepackage{bm}% bold math
\usepackage{mathtools}
\usepackage{xcolor}
\usepackage{layouts}
\usepackage{hyperref}
\usepackage{amssymb}
\usepackage{dsfont}

\usepackage{blindtext}

\newcommand{\tr}{\text{Tr}}
\newcommand{\state}[1]{|#1\rangle}
\newcommand{\conjstate}[1]{\langle #1|}
\newcommand{\pure}[1]{\state{#1}\conjstate{#1}}
\renewcommand{\d}{{\rm d}}
\newcommand{\1}{\mathds{1}}

\newcommand{\affilITP}{Institute for Theoretical Physics, ETH Z\"{u}rich, CH-8093 Z\"urich, Switzerland.}
\newcommand{\affilKON}{Department of Physics, University of Konstanz, 78464 Konstanz, Germany.}
\newcommand{\affilLAN}{Department of Physics, Lancaster University, Lancaster LA1 4YB, United Kingdom.}

\begin{document}

\preprint{APS/123-QED}

\title{Quantifying measurement-induced quantum-to-classical crossover using an open-system entanglement measure}

\author{Christian Carisch}\affiliation{\affilITP}
\author{Alessandro Romito}\affiliation{\affilLAN}
\author{Oded Zilberberg}\affiliation{\affilKON}

\date{\today}

\begin{abstract}
The evolution of a quantum system subject to measurements can be described by stochastic quantum trajectories of pure states. Instead, the ensemble average over trajectories is a mixed state evolving via a master equation.
Both descriptions lead to the same expectation values for linear observables.
Recently, there is growing interest in the average entanglement appearing during quantum trajectories.
The entanglement is a nonlinear observable that is sensitive to so-called  measurement-induced phase transitions, namely, transitions from a  system-size dependent phase to a quantum Zeno phase with area-law entanglement.
Intriguingly, the mixed steady-state description of these systems is insensitive to this phase transition. Together with the difficulty of quantifying the mixed state entanglement, this favors quantum trajectories for the description of the quantum measurement process.
Here, we study the entanglement of a single particle under continuous measurements (using the newly developed configuration coherence) in both the mixed state and the quantum trajectories descriptions.
In both descriptions, we find that the entanglement at intermediate time scales shows the same qualitative behavior as a function of the measurement strength.
The entanglement engenders a notion of coherence length, whose dependence on the measurement strength is explained by a cascade of underdamped-to-overdamped transitions.
This demonstrates that measurement-induced entanglement dynamics can be captured by mixed states. 
\end{abstract}

\maketitle

A quantum system is described by a wave function and, unlike its classical counterpart, can assume several states at once (superposition), where each state is associated with a certain (probability) amplitude. The time evolution of these amplitudes is governed by the famous Schr\"odinger wave equation~\cite{schroedinger_1926}. However, when we measure the particle in a specific classical state, the wavefunction's superposition must abruptly collapse with a state-dependent probability~\cite{Born_1926, von_neumann_1996,wiseman_milburn_2009}.
This stochastic process is incompatible with the deterministic Schr\"odinger equation. 
Over the years, various attempts have been made to integrate the measurement postulate into the framework of continuous wavefunction evolution by coupling the system to a detector~\cite{von_neumann_1996,wiseman_milburn_2009,gurvitz_1997,gurvitz_1998, gurvitz_et_al_2003}. In this case too, however, the quantum system effectively becomes open in presence of the out-of-equilibrium detectors, and measurement backaction on the system requires a statistical average over the quantum detector states. As a result, the wavefunction's time evolution under a sequence of measurements can be described by a \emph{quantum trajectory}~\cite{breuer_petruccione_2002, wiseman_milburn_2009}: the continuous evolution governed by the Schr\"odinger equation is interrupted by stochastic jumps whenever a measurement occurs.

Due to this emergent stochasticity, we can also consider the evolution of the average
probability density distribution of the measurement outcomes. 
This engenders a continuous evolution of the system's \emph{density matrix} using Lindblad's master equation~\cite{lindblad_1976, gorini_et_al_1976}.
Alternatively, the Lindblad master equation can be derived from the Schr\"odinger equation of the combined system and detector by tracing out the detector's degree of freedom in the limit of weak system-detector coupling and Markovian detector's dynamics~\cite{gurvitz_1997,breuer_petruccione_2002, wiseman_milburn_2009, Thomas2012, manzano_2020}.
Note that different assumptions on the detector and its coupling to the system lead to different types of master equations, including Nakajima-Zwanzig~\cite{nakajima_1958, zwanzig_1960}, Bloch-Redfield~\cite{wangsness_bloch_1953, bloch_1957, redfield_1965}, or the time-convolutionless master equations~\cite{tokuyama_mori_1975, tokuyama_mori_1976, breuer_et_al_2001, ferguson_et_al_2021}, can incorporate higher orders of system-detector coupling~\cite{zilberberg2014measuring,bischoff2015measurement,ferguson2020quantum}, and lead to exotic measurement protocols~\cite{zilberberg2011charge,zilberberg2013null,zilberberg2016many,zilberberg2019sensing}.
For the purpose of this work, we will remain within the Lindblad master equation framework.

\begin{figure*}[ht]
    \centering
    \includegraphics[width=\textwidth]{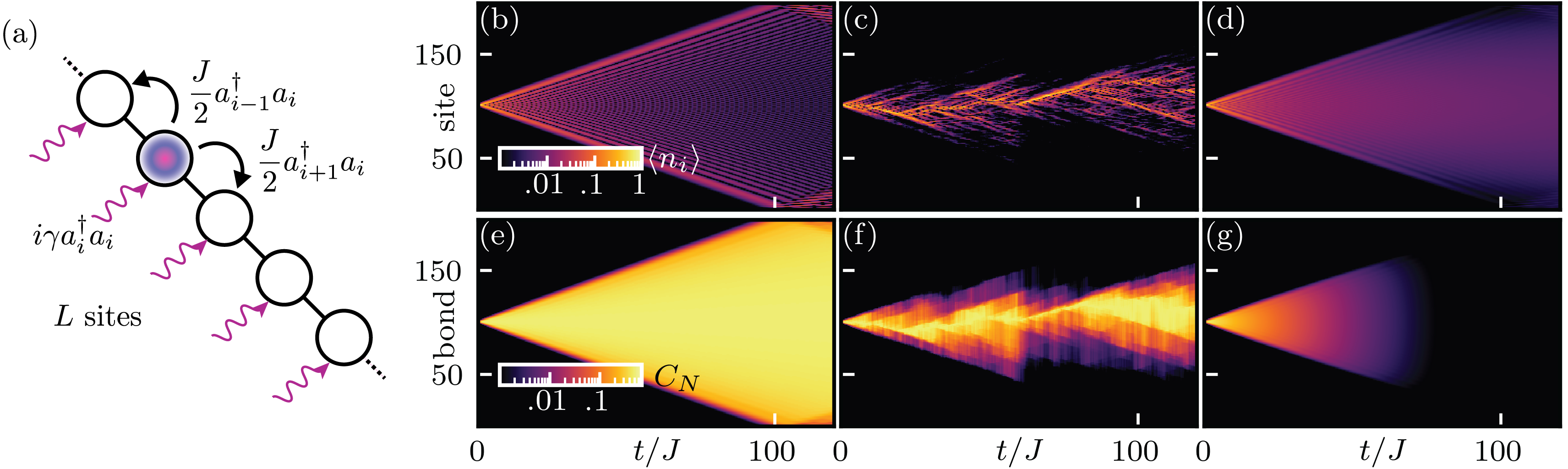}
    \caption{
    (a) Setup: spinless particles hopping with amplitude $J/2$ on a chain of length $L$ [cf.~Eq.~\eqref{eq: Hamiltonian}]. The particles are monitored with strength $\gamma$ [cf.~Eqs.~\eqref{eq: SSE} and~\eqref{eq: Lindblad}].
    (b)-(d) Evolution of local densities $\langle n_i\rangle$.
    (e)-(g) Evolution of the configuration coherence $C_N$.
    (b) and (e): Without monitoring ($\gamma=0$), the evolution is ballistic and supported by extensive entanglement.
    (c) and (f): Quantum random walk of a single quantum trajectory for measurement strength $\gamma=0.1$ [cf.~Eq.~\eqref{eq: SSE}].
    The ballistic evolution becomes interrupted by stochastic collapses.
    (d) and (g): Quantum-to-classical crossover of the mixed state evolution for measurement strength $\gamma=0.04$ [cf.~Eq.~\eqref{eq: Lindblad}; throughout this work using \textsc{QuTiP}~\cite{johansson_et_al_2013}].
    The evolution changes from ballistic at short times to diffusive at long times.
    }
    \label{fig: 1}
\end{figure*}

Recently, the equivalence between the quantum trajectory and Lindblad master equation descriptions has been challenged in the context of measurement-induced phase transitions. Here, the competition between the coherent evolution and the measurement collapse leads to a phase transition that is commonly quantified using an entanglement measure as an order parameter. 
Specifically, one observes a transition from a critical phase with system-size dependent entanglement for weak measurements to an area-law quantum Zeno phase for strong measurements~\cite{li_et_al_2018, szyniszewski_et_al_2019, skinner_et_al_2019, li_et_al_2019, szyniszewski_et_al_2020}, 
which has been reported in early experiments~\cite{noel_et_al_2022, koh_et_al_2022}.

%The measurement-induced entanglement phase transition has been discussed in the context of continuous measurements~\cite{szyniszewski_et_al_2020,alberton_et_al_2021,buchhold2021, biella_schiro_2021,turkeshi_et_al_2021, fava_et_al_2023}, non-hermitian dynamics~\cite{gopalakrishnan_gullans_2021,gal2022,kells2023,fleckenstein2022, turkeshi_2023}, purification transitions~\cite{gullans_huse_2020, bao_et_al_2020}, quantum error  correction~\cite{choi_et_al_2020, fan_et_al_2021, li_fisher_2021}, statistical mechanics~\cite{jian_et_al_2021, nahum_et_al_2021, lunt_et_al_2021, agrawal2022}, conformal field theory~\cite{li_et_al_2021}, Anderson localization~\cite{szyniszewski_et_al_2022, poepperl_et_al_2023}, and has been reported in experiments~\cite{noel_et_al_2022, koh_et_al_2022}.

Crucially, the order parameters used to quantify measurement-induced phase transitions are nonlinear functions of the density matrix, leading to a different outcome when averaging over sample paths or over the (density matrix) ensemble. 
Curiously, the mixed state described by the Lindlbad master equation shows no such phase transition because at long times, the disentangling measurements will always defeat the entangling effects~\cite{li_et_al_2018, skinner_et_al_2019, cao_et_al_2019, bao_et_al_2020}.
Moreover, while the entanglement of quantum trajectories can be efficiently measured by means of the entanglement entropy~\cite{nielsen_chuang_2010}, it is still notoriously difficult to extract the entanglement of the mixed state~\cite{gurvits_2003, gharibian_2008}.
This striking difference has sparked a  discussion about which of the quantum measurement descriptions is more revealing, with significant implications for a wide range of research fields, including quantum devices in the NISQ era~\cite{preskill_2018, brooks_2019} and quantum metrology~\cite{giovannetti_et_al_2011, pezze_et_al_2018, pirandola_et_al_2018}.

Here, we resolve the discrepancy between the two descriptions in capturing the measurement-induced entanglement dynamics.
To quantify entanglement in both the Lindblad and the quantum trajectory descriptions, we employ the recently developed configuration coherence~\cite{van_nieuwenburg_zilberberg_2018, carisch_zilberberg_2023}.
For simplicity, we study a single particle in the presence of local density measurements. 
The corresponding dynamics of monitored free fermions has been studied at trajectory levels~\cite{alberton_et_al_2021, buchhold2021, turkeshi_et_al_2021, turkeshi_et_al_2022, fava_et_al_2023, gal2022, kells_et_al_2023, szyniszewski_et_al_2022, poepperl_et_al_2023, coppola_et_al_2022, merritt_fidkowski_2023, poboiko_et_al_2023} showing the presence of a measurement-induced area-law phase.
In the Lindblad description the detector resembles a dephasing environment.
As the density matrix thermalizes in the long-time limit, we study instead the short- and intermediate-time behavior of the system.
Here, we find that the quantum trajectories and the mixed state show a qualitatively similar entanglement evolution, and are able to extract a notion of coherence length using both approaches.
We further show that this coherence length saturates for large values of the measurement strength.
The saturation of the coherence length can be understood as a cascade of underdamped-to-overdamped transitions in the Liouvillian eigenmodes~\cite{supmat}.
Our results enable the investigation of the measurement-induced entanglement phase transition in the context of mixed states.

We consider a spinless particle hopping on a 1D chain in the presence of local density measurements, see Fig.~\ref{fig: 1}(a).
The particle's free evolution is described by the Hamiltonian 
\begin{equation}
\label{eq: Hamiltonian}
    H~=\frac{J}{2}~\sum_{i=1}^{L-1} \left[ a_i^\dag a_{i+1}^{\phantom{\dagger}} + a_{i+1}^\dag a_i^{\phantom{\dagger}}\right]\, ,
\end{equation}
with $J$ the hopping amplitude, $a_i^\dag$ ($a_i$) the creation (annihilation) operator, and $L$ the chain's length. 
For convenience, we set $J=1$.
The spectrum $\epsilon(k)$ of the closed system~\eqref{eq: Hamiltonian} is associated with standing waves $\state{\tilde{k}_m}$ with group velocities $v_m \equiv\left(\partial \epsilon/\partial k\right)_{k=\tilde{k}_m}= \sin(\pi m/(L+1))$, $m~=~1,\dots,L$.
In the following, we inject the particle into the center of the chain, $\state{\psi(t=0)}=\state{L/2}\coloneqq a^\dag_{L/2}\state{0}$, where $\state{0}$ is the vacuum state.
Such a localized particle overlaps with all the eigenmodes simultaneously, resulting in a ballistic quantum random walk~\cite{kempe_2003}.
Its characteristic density envelope evolves linearly with velocity $v_{L/2}^{\phantom{\dagger}}\approx 1$ of the fastest eigenmode~\cite{schoenhammer_2019}, see Fig.~\ref{fig: 1}(b).

In general, measurements of the particle will modify the coherent ballistic evolution~\cite{esposito_gaspard_2005, basko_et_al_2006, amir_et_al_2009, znidaric_2010}. As a simple model of quantum measurement, we consider that the chain's sites are capacitively coupled to independent detectors, with coupling strength $\gamma\geq 0$, see Fig.~\ref{fig: 1}(a).
Specifically, the detectors monitor the state's local densities $\langle n_i \rangle~\equiv~\conjstate{\psi} a_i^\dag a_i^{\phantom{\dagger}} \state{\psi}$.
The evolution of the system using a quantum trajectory description follows the stochastic Schr\"odinger equation (SSE)~\cite{wiseman_milburn_2009, cao_et_al_2019, turkeshi_et_al_2021}
\begin{align}
    \label{eq: SSE}
    &\d \state{\psi} = -i H \d t \state{\psi}  \\ &+ \sum_{i=1}^L \Big(\sqrt{\gamma} [n_i - \langle n_i \rangle  ]\d W^i_t -\frac{\gamma}{2}[n_i - \langle n_i \rangle]^2 \d t \Big) \state{\psi} \nonumber \, , 
\end{align}
where $\gamma$ is the coupling rate to the detectors (the measurement strength), and $\d W^i_t$ are Wiener increments with $\langle \d W^i_t \d W^j_{t'}\rangle =  \delta_{i,j}\delta_{t,t'} \d t$.
The particle's time evolution follows a stochastic sample path in space, a.k.a. quantum trajectory, see Fig.~\ref{fig: 1}(c).
Whenever a measurement occurs, the ballistic evolution of the trajectory is interrupted by a collapse (quantum jump). The trajectory describes one possible sequence of measurement events and outcomes.

To account for all possible measurement sequences, one commonly samples the SSE for a large number $M\gg 1$ of quantum trajectories with equal initial conditions, i.e., $\state{\psi_i(t~=~0)}=\state{L/2}$, $i=1,\dots,M$.
The average probability distribution of possible outcomes leads to a  \emph{mixed state} that is described by the density matrix $\rho~=~\overline{\pure{\psi_i}}~=~\frac{1}{M} \sum_{i=1}^M\pure{\psi_i}$.
Alternatively, instead of an average over quantum trajectories, the evolution of the density matrix $\rho$ itself can be described by the Lindblad equation~\cite{wiseman_milburn_2009, cao_et_al_2019}
\begin{equation}
    \label{eq: Lindblad}
    \partial_t \rho = -i [H, \rho] + \gamma \sum_{i=1}^L \left( n_i \rho n_i - \frac{1}{2}\{ n_i, \rho\}\right)\, .
\end{equation}
For a finite measurement strength $\gamma > 0$, the dynamics of the density matrix of the initially localized particle changes from ballistic at times $t\lesssim 1/\gamma$ to diffusive at $t\gtrsim 1/\gamma$~\cite{amir_et_al_2009}, see Fig.~\ref{fig: 1}(d). This transition is akin to a quantum-to-classical crossover. 
Interestingly, this crossover is not visible on the single trajectory level. In the following, we will characterize the crossover by comparing the entanglement in the system in both the quantum trajectory and the density matrix descriptions.
Because entanglement is a nonlinear quantity (order parameter), we expect different results for the mixed state and the trajectory-averaged entanglement~\cite{skinner_et_al_2019, cao_et_al_2019}. 
This has favoured the use of trajectories rather than the Lindblad evolution to characterize measurement-induced entanglement dynamics~\cite{skinner_et_al_2019, cao_et_al_2019, bao_et_al_2020}.

\begin{figure}[t!]
    \centering
    \includegraphics[width=\columnwidth]{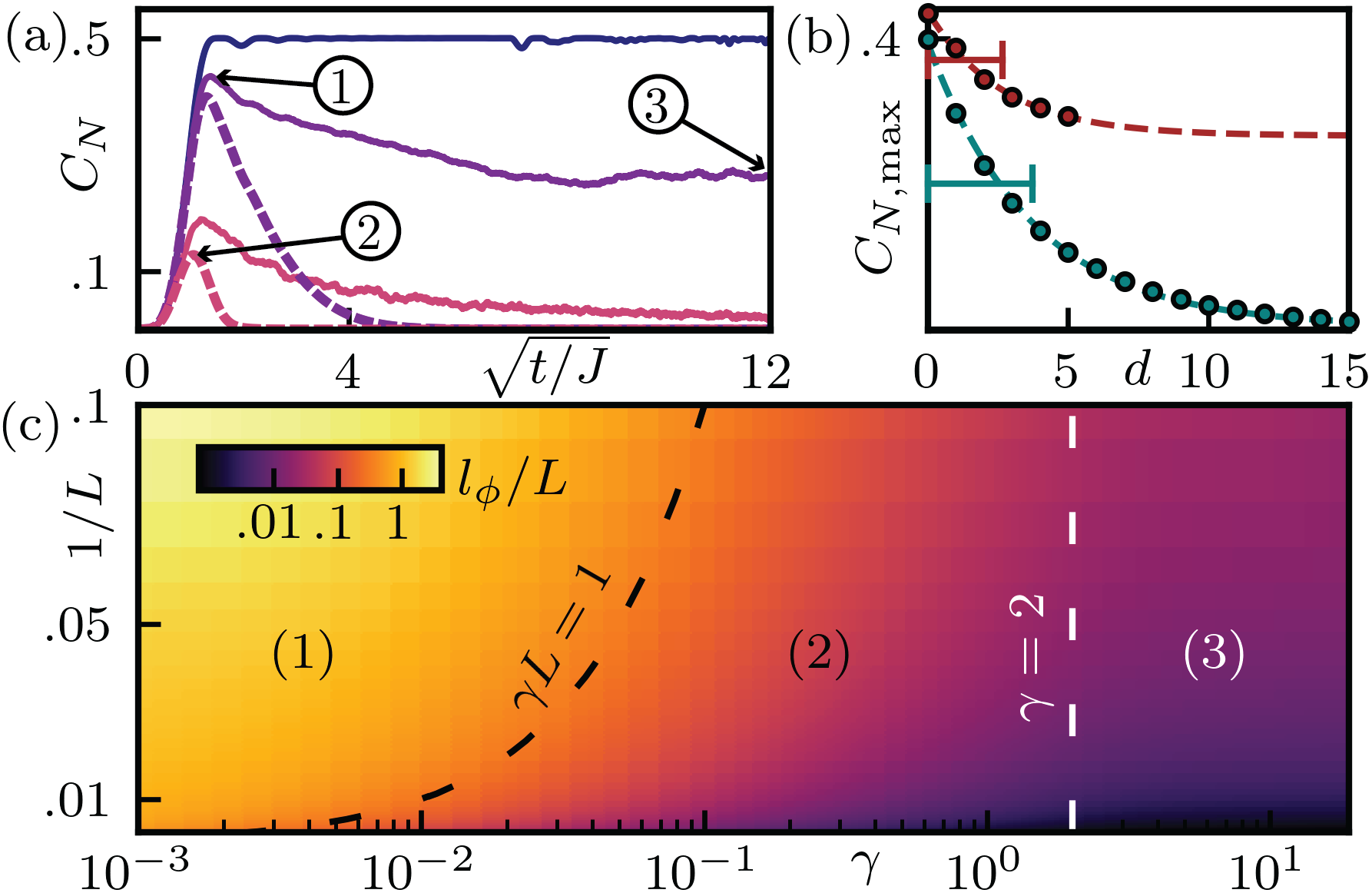}
    \caption{
    (a) The configuration coherence at the middle bond as a function of time for quantum trajectories (solid lines) and the mixed state (dashed lines).
    Chain length is $L=50$, trajectory values are averaged over $10^3$ runs, and measurement strenghts are $\gamma=0,\, 0.1,\,1$ (blue, purple, pink).
    For $\gamma=0$, the descriptions are equivalent. 
    For finite $\gamma$, both descriptions show a maximum at an intermediate time [markers \raisebox{.5pt}{\textcircled{\raisebox{-.9pt} {1}}}, \raisebox{.5pt}{\textcircled{\raisebox{-.9pt} {2}}}] followed by a decay to a finite [marker \raisebox{.5pt}{\textcircled{\raisebox{-.9pt} {3}}}] (zero) saturation value at long times for the trajectory (mixed state) description.
    (b) The maximal configuration coherence $C_{N,{\rm max}}$ as a function of the distance $d$ from the injection point of the single particle for measurement strength $\gamma=0.1$ [cf.~Eq.~\eqref{eq: exponential fit}]. An exponential fit (dashed lines) determines the coherence length $l_\phi$ for the mixed state (green) and $\lambda_\phi$ for trajectories (brown) marked by horizontal bars.
    (c) Phase diagram of the normalized single particle coherence length $\l_\phi/L$ of the mixed state.
    We find 3 phases: (1) for small measurement strengths $\gamma \lesssim 1/L$, the particle coherently explores the full chain; (2) for intermediate measurement strengths $1/L\lesssim \gamma \lesssim 2$, the coherence length is finite and depends on the measurement strength, $l_\phi = l_\phi(\gamma)$; (3) for large measurement strengths $\gamma \gtrsim 2$, the coherence length saturates to $l_{\phi, \infty}\approx 0.3$.
    }
    \label{fig: 2}
\end{figure}

For a 1D chain, entanglement describes the quantum correlations with respect to a cut at a bond $b$.
To quantify entanglement, we employ the configuration coherence~\cite{van_nieuwenburg_zilberberg_2018, carisch_zilberberg_2023}
\begin{equation}
\label{eq: entanglement}
    C_N(\rho, b) = 2\sum_{\mathclap{\substack{i=1,\dots,b \\ j=b+1,\dots, L }}}|\conjstate{i} \rho \state{j}|^2\, , 
\end{equation}
where $\state{j}=a_j^\dag \state{0}$. For pure states, the configuration coherence is as $C_N(\state{\psi}, b)\coloneqq C_N(\pure{\psi}, b)$. The configuration coherence is a convex entanglement measure for mixed states under certain conditions, e.g., with a fixed number of particles subject to Lindblad evolution with Hermitian jump operators.
For our case study, both the SSE~\eqref{eq: SSE} as well as the Linbdlad equation~\eqref{eq: Lindblad} fall under this category.
For a single particle, the configuration coherence is related to the negativity $\mathcal{N}(\rho)$~\cite{vidal_werner_2002}.
In fact, $\sqrt{C_N(\rho)/2}=\mathcal{N}(\rho) \equiv (||\rho^{\rm T_B}||_1 - 1)/2$, where $\rho^{\rm T_B}$ is the partial transpose and $|| \cdot ||_1$ denotes the trace norm.
In Figs.~\ref{fig: 1}(e)-(g), we plot the configuration coherence at each bond for the single particle evolutions discussed so far, cf.~Figs.~\ref{fig: 1}(b)-(d).
Without measurements ($\gamma=0$), the particle ballistically evolves into a superposition over all sites, leading to the expansion of orbital entanglement across the chain, see Fig.~\ref{fig: 1}(e). 
For finite $\gamma$, the quantum trajectory exhibits entanglement expansion with intermittent collapses, see Fig.~\ref{fig: 1}(f). Crucially, we observe finite entanglement at long times $t\gg 1/\gamma$ for $\gamma>0$. Conversely, the mixed state's entanglement vanishes around $t\approx 1/\gamma$, justifying the quantum-to-classical crossover labeling, see Fig.~\ref{fig: 1}(g).
As expected from Lindblad evolution, the density matrix $\rho$ evolves into a non-entangled infinite temperature state, $\rho(t \to \infty) = \1/L$. 
The time $t\approx 1/\gamma$ can be understood as the system's dephasing or Thouless' time~\cite{edwards_thouless_1972, akkermans_montambaux_2007}. 
Note that the vanishing entanglement at long times is the second motivation to favor quantum trajectories when studying measurement-induced phase transitions~\cite{li_et_al_2018, cao_et_al_2019, bao_et_al_2020}.

Now, we consider the time evolution of the configuration coherence and arrive at a definition for a $\gamma$-dependent coherence length. The coherence (Thouless) length describes the length scale over which the particle can evolve ballistically before the quantum-to-classical crossover turns its motion diffusive. 
This is reminiscent to defining an entanglement-based order parameter for describing the physics of our system, cf.~Refs.~\cite{bayat_et_al_2014,iqbal_schuch_2021, stocker_et_al_2022}.
Again, we inject the single particle at the center of the chain.
First, we consider the average configuration coherence $\overline{C_N(\state{\psi_i}, b)}$ over the quantum trajectories at any bond, $b=1,\dots, L-1$ [see Fig.~\ref{fig: 2}(a)]:
after it assumes a maximal value $\overline{C_N(\state{\psi_i}, b)}_{\rm max}$ at intermediate times, it saturates to a $\gamma$-dependent finite value, $\overline{C_N(\state{\psi_i}, b)}_{\infty}\,$, for $t\to \infty$.
For each bond $b$ that satisfies $\overline{C_N(\state{\psi_i}, b)}_{\rm max} > \overline{C_N(\state{\psi_i}, b)}_{\infty}$, we plot the maximal configuration coherence as a function of the distance $d=|b - L/2|$ from the injection point, see Fig.~\ref{fig: 2}(b).
We find an exponential decay dependence,
\begin{equation}
\label{eq: exponential fit}
    \overline{C_N(\state{\psi_i}, b)}_{\rm max} \propto \exp(-d/\lambda_\phi)\, .
\end{equation}
We use this decay to define the particle's coherence length $\lambda_\phi = \lambda_\phi(\gamma)$.
Second, we consider the configuration coherence evolution of the mixed state.
The mixed state exhibits a maximal value at a similar intermediate time, before it decays and vanishes, see Fig.~\ref{fig: 2}(a). Again, we can extract the coherence length $l_\phi$ of the single particle as the length scale of the configuration coherence's exponential decay as a function of the distance from the injection point, $C_N(\rho, b)_{\max} \propto \exp(-d/l_\phi)$, see Fig.~\ref{fig: 2}(b).
Next, we analyze in more detail how the mixed state's coherence length depends on the measurement strength $\gamma$.

We use the mixed state entanglement analysis approach. The coherence length $l_\phi$ shows three distinct regimes as a function of the measurement strength $\gamma$, see Fig.~\ref{fig: 2}(c):
(1) For weak measurements $\gamma \lesssim 1/L$, the whole chain is explored coherently, and $\l_\phi \gtrsim L$ is independent of the measurement strength. This is the regime of mesoscopic physics~\cite{imry_2002, akkermans_montambaux_2007}.
(2) For intermediate measurement strengths $2\lesssim \gamma \lesssim 1/L$, the particle coherently explores a region of width $l_\phi \leq L$, beyond which it evolves diffusively.
(3) For strong measurements $\gamma \gtrsim 2$, the coherence length saturates to $l_{\phi, \infty} \approx 0.3$. As such, scaling the system's size leads to three qualitatively different regions/phases.
We employ the same analysis for the quantum trajectories and find a qualitatively similar behavior of the coherence length $\lambda_\phi$, see~\cite{supmat}:
the crossover from (2) to (3) happens for $\gamma \approx 1.3$ and the saturating value is $\lambda_{\phi, \infty} \approx 1.7$.
The observation of these regimes in the coherence length of the single-particle mixed state is the main result of our work. 

\begin{figure}[t!]
    \centering
    \includegraphics[width=\columnwidth]{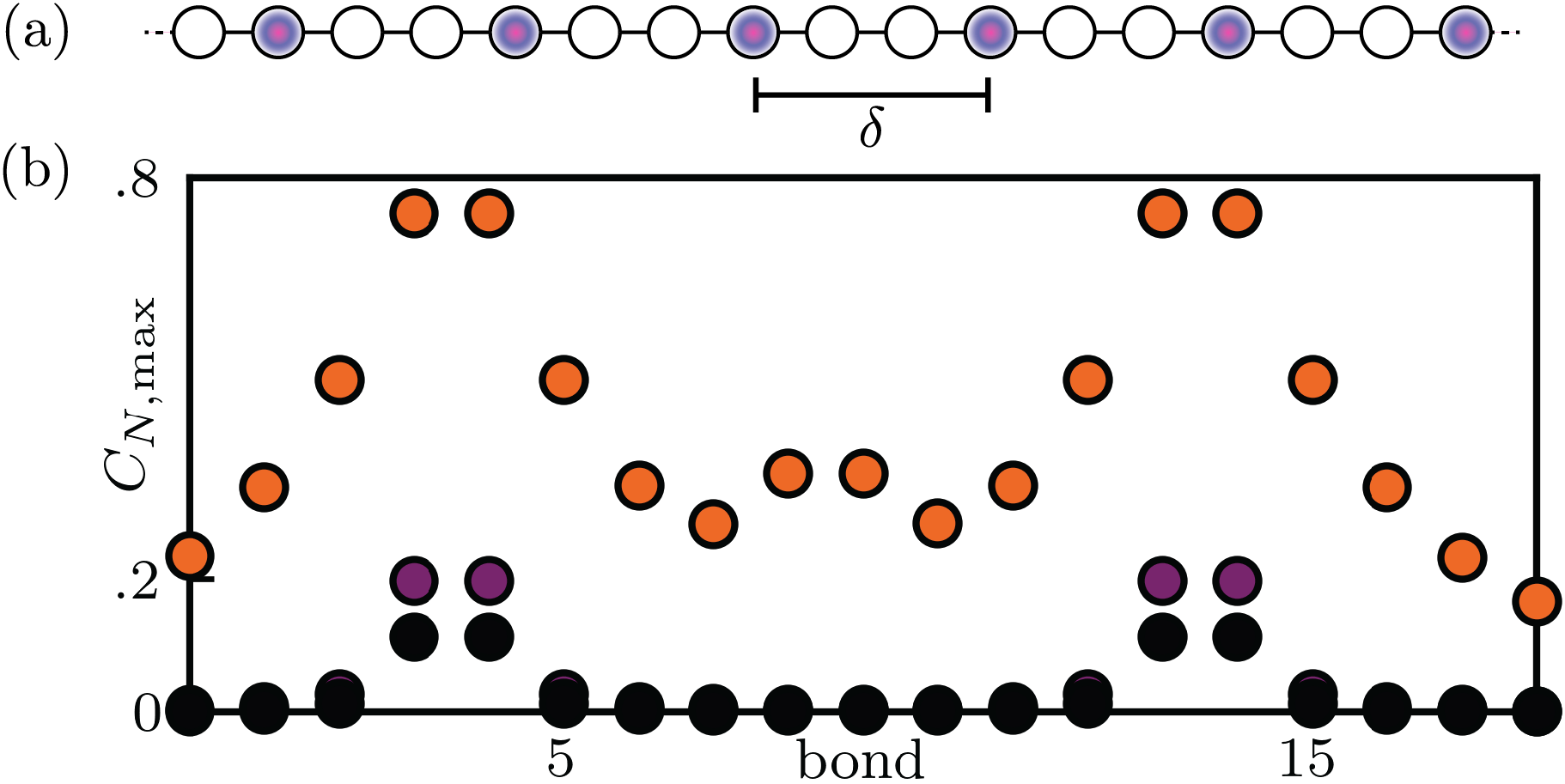}
    \caption{
    (a) Chain with many particles homogeneously distributed with interparticle distance $\delta$.
    (b) Maximal configuration coherence for two particles on $L=20$ sites, injected at sites 5 and 15, for measurement strengths $\gamma=0.1, \ 1,\ 2$ (orange, purple, black).
    For $\gamma=0.1$, two-particle entanglement  between the two injection points builds up, because the particles coherently interact with each other.
    For large measurement strengths $\gamma =[1, \ 2]$, the maximal configuration coherence is the sum over the single-particle contributions.
    Our numerical implementation harnesses a matrix product density operator~\cite{verstraete_et_al_2004} representation of the mixed state.
    %Heuristic sketch of the logarithmic configuration coherence at the half chain as a function of the interparticle distance $\delta$ for two different values of the measurement strength $\gamma$. For a dilute system, the coherence length is shorter than $\delta$, and the peak entanglement is the same for both measurement strengths. For a dense system, the particles start to coherently overlap and the long coherence length for $\gamma =0.2$ leads to increasing entanglement. On the other hand, for $\gamma=2$, the coherence length is shorter than the lattice constant, and the peak entanglement remains the same as for the dilute system.
    }
    \label{fig: 3}
\end{figure}

The metallic-to-diffusive transition in the dynamics of our system has been studied in various setups~\cite{esposito_gaspard_2005, basko_et_al_2006, amir_et_al_2009, znidaric_2010}.
The crossover from (1) to (3) can be understood as a sequence of underdamped-to-overdamped transitions of the Hamiltonian's plane wave eigenmodes~\eqref{eq: Hamiltonian}.
Specifically, each mode is damped by the measurement strength $\gamma$.
With increasing $\gamma$, the modes become overdamped, starting with the slowest of the modes.
At $\gamma \approx 2$, the fastest mode becomes overdamped, and the particle enters a regime where its coherent evolution is exponentially suppressed.
As such, this transition does not depend on the choice of the initial state and can also be seen in the Liouvillian spectrum of the system, and obtained analytically in the two-sites case~\cite{supmat}.
Note that such underdamped-to-overdamped transitions can also be observed in the Spin-Boson model~\cite{leggett_et_al_1987} or in double quantum dots under dephasing~\cite{gurvitz_1997}. 
It appears that such physics underpins the measurement-induced phase transition so that the latter lends an entanglement-based order parameter for the characterization of these effects~\cite {li_et_al_2018, skinner_et_al_2019, li_et_al_2019, biella_schiro_2021}. As we find here, such characterization can be accomplished not only in quantum trajectories but also from the mixed state evolution.

The question remains of how our single-particle toy model~\eqref{eq: Lindblad} generalizes to many-body systems.
For dilute systems, we postulate that the average inter-particle distance $\delta$ will replace the system size $L$ as the relevant length scale, see Fig.~\ref{fig: 3}(a). 
In the presence of strong measurements ($l_\phi < \delta/2$), the particles will not coherently experience each other.
The many-body entanglement will then separate into a sum of single-particle contributions. 
The onset of many-body entanglement happens when the coherence length becomes of the order of half the inter-particle distance, $l_\phi \approx \delta/2$.
In Fig.~\ref{fig: 3}(b), we show such two-particle entanglement as a local maximum between the two particles that cannot be described as a sum of single particle contributions~\cite{lukin2019probing,kaplan2020many,wu2020prethermal}.
%In line with the discussion of measurement-induced entanglement phase transitions, we expect a critical value of the measurement strength below which the particles can coherently interact over a long range.
For $l_\phi \ll \delta/2$, the entanglement will not depend on the chain length $L$ or the inter-particle distance $\delta$.
For $l_\phi \gtrsim \delta/2$, multiple particles can contribute to the configuration coherence and we, therefore, expect the entanglement to depend on the inter-particle distance $\delta$.
In light of the quantitative difference in the coherence lengths of the trajectories and the mixed state, we expect the onset of many-body entanglement at different values of the measurement strength $\gamma$.

%and we can approximate the many-body state of $N$ particles by a product of single particle density matrices,  $\rho_N \approx \bigotimes_{i=1}^N \rho_i$.
%Then, we can assess the many-body entanglement using the logarithmic negativity~\cite{vidal_werner_2002, plenio_2005}, which in terms of the configuration coherence is given as $E_{\N}(\rho)=\log_2(\sqrt{2C_N(\rho)} + 1)$.
%Crucially, the logarithmic negativity is an additive entanglement measure, and therefore $E_{\N}(\rho_N)=\sum_i E_{\N}(\rho_i)$. Hence, as soon as the inter-particle distance becomes comparable to the coherence length, the particles will coherently interact and change the entanglement behavior, as sketched in Fig.~\ref{fig: 3}(b).
%Note that at this point, the many-body state cannot be described by a product state anymore, and the total entanglement will not be a simple sum of the single-particle entanglement contributions~\cite{lukin2019probing,kaplan2020many,wu2020prethermal}.

We have analyzed the quantum measurement of a single particle using both quantum trajectories and a mixed-state Lindblad description.
For both descriptions, we employed the recently developed configuration coherence as an entanglement measure~\cite{van_nieuwenburg_zilberberg_2018,carisch_zilberberg_2023}.
At first sight, the entanglement behaviour of the trajectories is opposite to that of the mixed state because the former remains finite at long times.
At intermediate times, however, we extracted a coherence length from the entanglement and showed that it behaves qualitatively the same for both descriptions.
Moreover, we found that the coherence length saturates at a finite value for large measurement strengths, namely when the fastest Liouvillian eigenmode becomes overdamped.
Besides new insights into the quantum-to-classical crossover in terms of entanglement, our results provide evidence that the master equation can capture the measurement-induced entanglement dynamics of monitored systems. This observation unveils the underlying stochastic physics that measurements impart on the system, and their commonly-observed manifestation in diffusive dynamics.  In future work, we will extend the discussion to the many-body case.

\emph{Acknowledgments.} We thank C.~Leung for help with the quantum trajectory code and M.~H.~Fischer and J.~del~Pino for fruitful discussions. The authors acknowledge financial support by ETH Research Grant ETH-51 201-1 and the Deutsche Forschungsgemeinschaft (DFG) - project number 449653034.

\bibliography{library}

\newpage
	\cleardoublepage
	\setcounter{figure}{0}

	{\onecolumngrid
		\begin{center}
			\textbf{\normalsize Supplemental Material for}\\
			\vspace{3mm}
			\textbf{\large Quantifying measurement-induced quantum-to-classical crossover using an open-system entanglement measure}
			\vspace{4mm}
			
			{ Christian Carisch$^{1}$, Alessandro Romito$^{2}$, and Oded Zilberberg$^{3}$}\\
			\vspace{1mm}
			\textit{\small $^{1}$Institute for Theoretical Physics, ETH Z\"urich, 8093 Z\"urich, Switzerland\\
			}
            \textit{\small $^{2}$Department of Physics, Lancaster University, Lancaster LA1 4YB, United Kingdom}\\
			\textit{\small $^{3}$Department of Physics, University of Konstanz, D-78457 Konstanz, Germany}
			
			\vspace{5mm}
	\end{center}}
	
	\twocolumngrid
	
	\setcounter{equation}{0}
	\setcounter{section}{0}
	\setcounter{figure}{0}
	\setcounter{table}{0}
	\setcounter{page}{1}
	\makeatletter
	\renewcommand{\bibnumfmt}[1]{[#1]}
	\renewcommand{\citenumfont}[1]{#1}
	
	\setcounter{enumi}{1}
	\renewcommand{\theequation}{S\arabic{equation}}
	\renewcommand{\thesection}{S\arabic{section}}
	\renewcommand{\thetable}{S\arabic{table}}
	\renewcommand{\thefigure}{S\arabic{figure}}

	\section{1. Partial scaling collapse of the coherence length of the mixed state}
In the main text, we extract the coherence length $l_\phi$ using the mixed-state evolution of a single particle under dephasing.
The coherence length is a function of two parameters, the chain's length $L$ and the measurement strength $\gamma$.
Here, we show in detail the dependence of the coherence length on these parameters.
We find two distinct scaling regimes.

\begin{figure}[t!]
    \centering
    \includegraphics[width=\columnwidth]{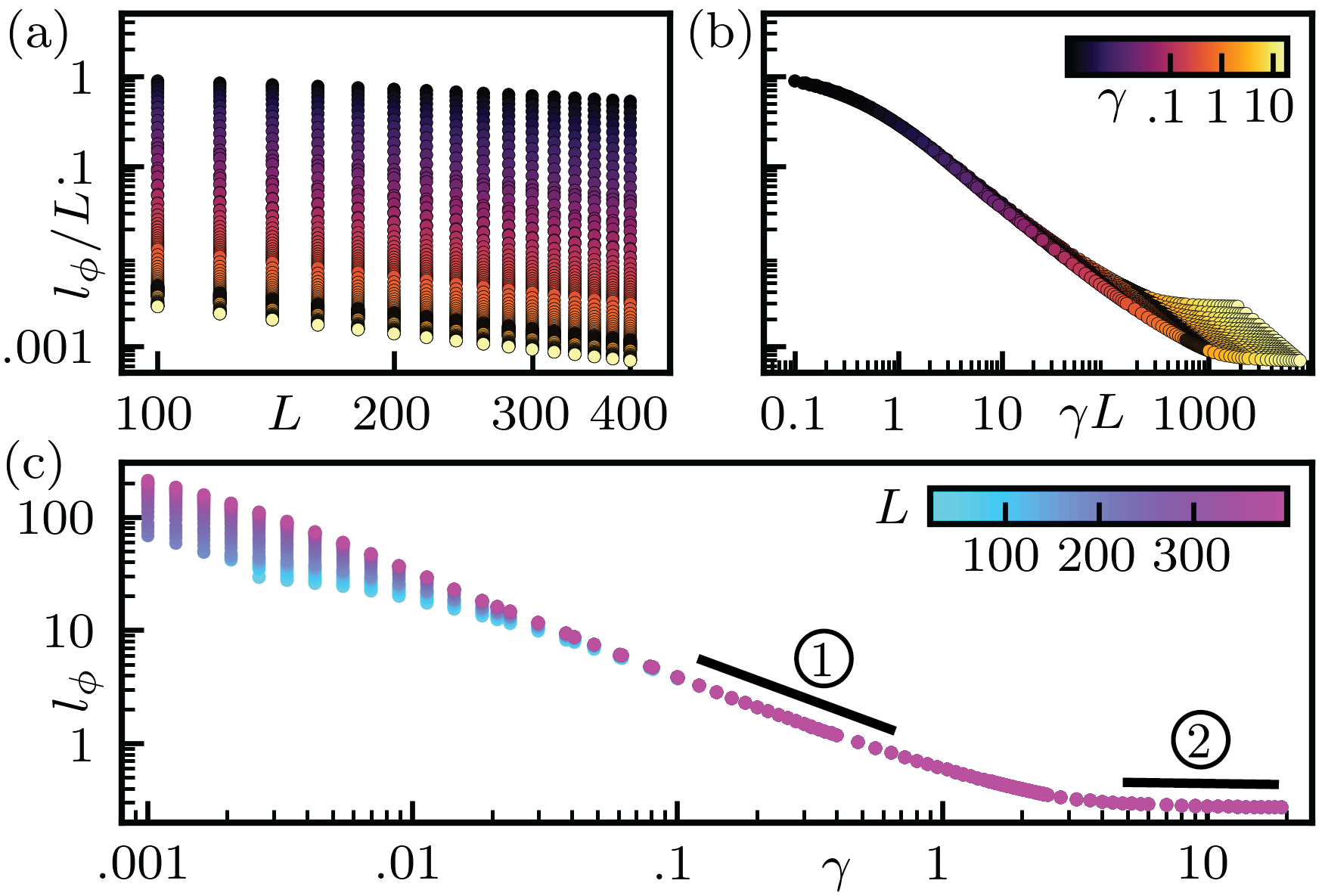}
    \caption{
    Scaling collapse of the normalized coherence length.
    (a) The normalized coherence length $l_\phi/L$ as a function of the chain length $L$ for different values of the measurement strength $\gamma$.
    (b) Same data as (a) but with a rescaled x-axis $\gamma L$. For small values of $\gamma L$, the normalized coherence length is a function of this parameter only (all data points collapse on a single line).
    For larger values, the curves separate.
    (c) The coherence length $l_\phi$ as a function of the measurement strength $\gamma$ for different chain lengths.
    For small measurement strengths, we observe finite-size effects. For large measurement strengths, the curves collapse, and we find a universal behavior of the coherence length, independent of the chain's length $L$.
    Marker \raisebox{.5pt}{\textcircled{\raisebox{-.9pt} {1}}}: regime with power-law scaling of the coherence length, $l_\phi \sim a/\gamma^b$.
    Marker \raisebox{.5pt}{\textcircled{\raisebox{-.9pt} {2}}}: saturated regime, $l_\phi \approx l_{\phi, \infty}$.
    }
    \label{fig: scaling collapse}
\end{figure}

1. \emph{Weak measurement strengths.} The first regime is obtained using a scaling collapse of the normalized coherence length $l_\phi/L$ for weak measurement strengths $\gamma$.
In Fig.~\ref{fig: scaling collapse}(a), we plot the normalized coherence length for various chain lengths $L$ and measurement strengths [$\gamma$ increases from top to bottom for each column, cf.~color bar in Fig.~\ref{fig: scaling collapse}(b)].
A rescaling of the x-axis to the dimensionless parameter $\gamma L$ collapses all curves onto a single one for small values of the parameter, see Fig.~\ref{fig: scaling collapse}(b).
This means that the scaling behavior of the normalized coherence length can be described by a single parameter $\gamma L$ as long as the measurement strength is weak enough for the particle to coherently explore the full extent of the chain.

2. \emph{Large measurement strengths.} For large values $\gamma L \gtrsim 10$, the different curves separate, see Fig.~\ref{fig: scaling collapse}(b).
This is because the measurement strength is high enough to localize the coherence of the particle far enough away from the chain's ends.
Indeed, we find that the coherence length $l_\phi$ develops a universal behavior as a function of the measurement strength $\gamma$ if the latter is strong enough.
This is evident from Fig.~\ref{fig: scaling collapse}(c), where the curves for different chain lengths $L$ collapse onto a single line for $\gamma \gtrsim 0.05$.
The collapsing value corresponds to $\gamma \approx 1/L_{\rm min}$, with the smallest sampled chain length $L_{\rm min}=20$.
Note that the smallest chain with a notion of a coherence length has length 2, so we find a system-size independent scaling of the coherence length for $\gamma\gtrsim 0.5$.

\section{2. Extraction of the saturating measurement strength}

In the main text, we found that the coherence length $l_\phi$ saturates for large values of the measurement strength $\gamma$.
The saturation can also be observed in Fig.~\ref{fig: scaling collapse}(c).
Here, we explain how we estimate the measurement strength at which the saturation occurs.

In Fig.~\ref{fig: scaling collapse}(c), we identify two  different functional dependencies of $l_\phi(\gamma)$: \raisebox{.5pt}{\textcircled{\raisebox{-.9pt} {1}}} for $1\gtrsim \gamma \gtrsim0.1$, we find a linear dependence in the loglog plot, i.e., $l_\phi(\gamma) \sim a/\gamma^b$ for constants $a,\ b\geq 0$;
\raisebox{.5pt}{\textcircled{\raisebox{-.9pt} {2}}} for large $\gamma$, the coherence length saturates to $l_\phi(\gamma)=l_{\phi, \infty}$.
We obtain an estimate of the value $\gamma$ that separates the two regimes by intersecting their lines,
\begin{equation}
\label{eq: saturating gamma}
    \frac{a}{\gamma^b} = l_{\phi, \infty} \quad \to \quad \gamma = \left(\frac{a}{l_{\phi, \infty}}\right)^\frac{1}{b}\, .
\end{equation}
A numerical fit to the data gives $a\approx0.5$ and $b\approx0.8$.
We estimate the saturation value $l_{\phi, \infty}\approx 0.3$ as the coherence length of longest chain, $L=400$, with highest measurement strength, $\gamma = 19$.
Plugging the values into~\eqref{eq: saturating gamma}, we obtain that the coherence length saturates around $\gamma \approx 1.9$.

\section{3. Coherence length of quantum trajectories}
\begin{figure}[t!]
    \centering 
    \includegraphics[width=\columnwidth]{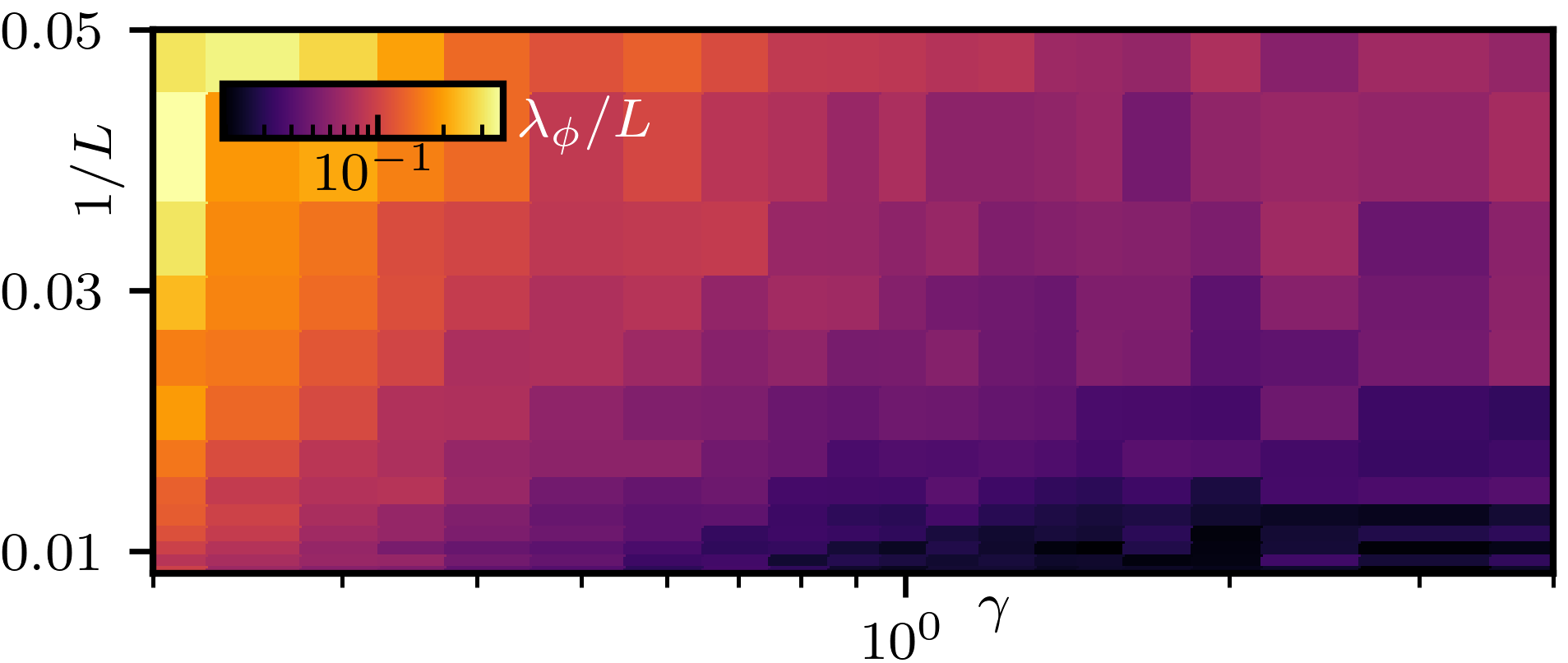}
    \caption{Phase diagram of the coherence length $\lambda_\phi$ extracted from an average over quantum trajectories for different chain lengths $L$ as a function of the measurement strength $\gamma$. We average over $400-1000$ trajectories for every data point. Numerical details: timestep $dt={\rm min}[0.0.5,  \ 0.05/\gamma]$, end time $t_{\rm max}={\rm max}[2L, \ 10/\gamma]$.
    }
    \label{fig: phase diagram qt}
\end{figure}

In this section, we discuss the behavior of the coherence length $\lambda_\phi$ of the quantum trajectories as a function of the measurement strength $\gamma$.
We show the phase diagram of the renormalized coherence length $\lambda_\phi/L$ in Fig.~\ref{fig: phase diagram qt}.
For small measurement strengths and system sizes, the coherence length is of the order of the system size (top left corner of the phase diagram).
For larger measurement strengths, the coherence length decreases with increasing measurement strength, $\lambda_\phi = \lambda_\phi(\gamma)$.
We observe coherence length saturation for large values of $\gamma$, in qualitative agreement with the coherence length saturation for mixed states. 

\begin{figure}[t!]
    \centering 
    \includegraphics[width=\columnwidth]{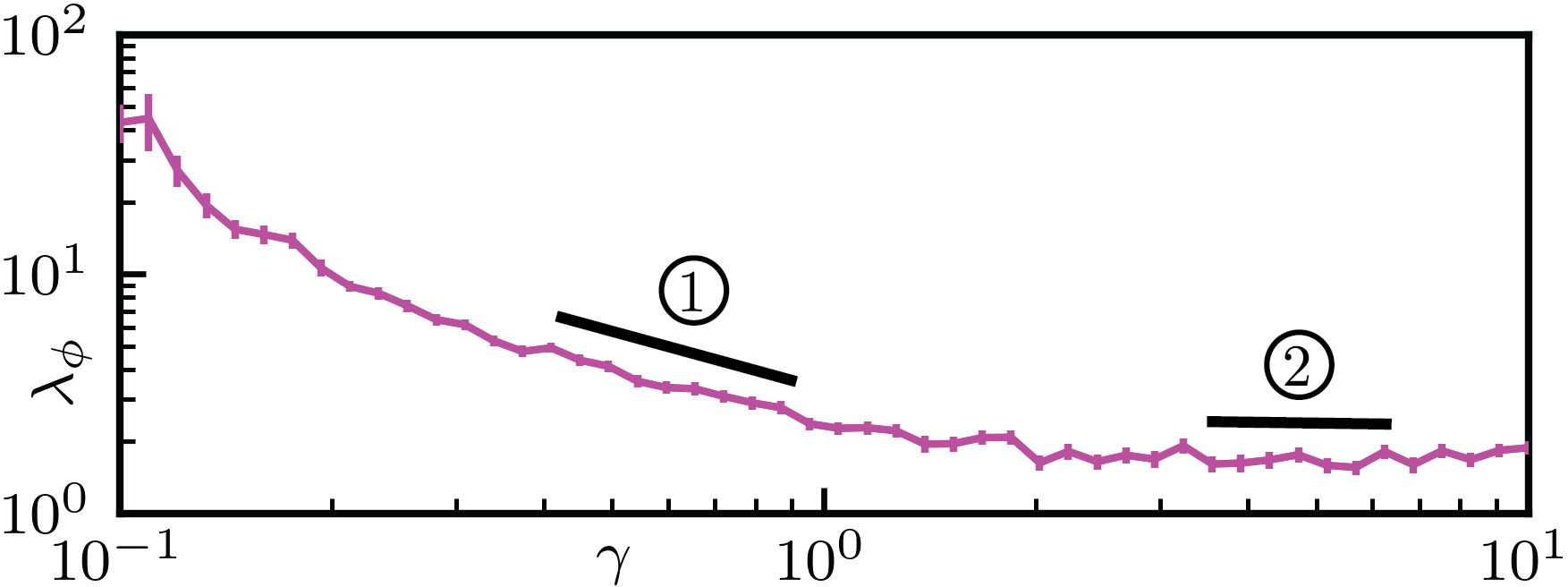}
    \caption{Coherence length $\lambda_\phi$ extracted from an average over quantum trajectories for chain length $L=100$ as a function of the measurement strength $\gamma$. We average over $800-3200$ trajectories for every data point (more sampling for higher measurement strengths). The coherence length saturates for $\gamma \gtrsim 1.3$.
    Marker \raisebox{.5pt}{\textcircled{\raisebox{-.9pt} {1}}}: regime with power-law scaling of the coherence length, $\lambda_\phi \sim a/\gamma^b$.
    Marker \raisebox{.5pt}{\textcircled{\raisebox{-.9pt} {2}}}: saturated regime, $\lambda_\phi \approx \lambda_{\phi, \infty}$.
    }
    \label{fig: saturation of coherence length}
\end{figure}

Fig.~\ref{fig: saturation of coherence length} shows the saturation of the coherence length $\lambda_\phi$ for $L=100$.
As for the mixed state in Fig.~\ref{fig: scaling collapse}(c), we find two scaling regimes, which we use to extract the measurement strength at which the coherence length saturates:
{\textcircled{\raisebox{-.9pt} {1}}} for $1\gtrsim \gamma \gtrsim0.2$, we find a linear dependence in the loglog plot, i.e., $\lambda_\phi(\gamma) \sim a/\gamma^b$ for constants $a,\ b\geq 0$;
\raisebox{.5pt}{\textcircled{\raisebox{-.9pt} {2}}} for large $\gamma$, the coherence length saturates to $\lambda_\phi(\gamma)=\lambda_{\phi, \infty}$.
In line with the mixed state, we use Eq.~\eqref{eq: saturating gamma}
to determine the saturating measurement strengths.
We find $a\approx 2.2$ and $b\approx 0.9$ using a linear fit for $0.8>\gamma > 0.2$.
For the saturated coherence length we find $\lambda_{\phi, \infty}\approx 1.7$ by averaging $\lambda_\phi$ for $\gamma > 2$ (we average due to statistical errors for each single data point).
Note the quantitative difference to the mixed state saturated coherence length, $l_{\phi, \infty} \approx 0.3$.
Using these numerical values, we find that the quantum trajectory coherence length saturates for $\gamma \gtrsim 1.3$.

%We show the result in Fig.~\ref{fig: saturation of coherence length} for different values of the chain length $L$.
%Interestingly, the coherence length first increases with increasing measurement strength $\gamma$, but the errors of the extracted exponential fit are relatively large.
%The quantum trajectories display a variant of Kramer's turnover~\cite{kramers_1940, gurvitz_1997}:
%a quantum particle can overcome a potential barrier either via tunneling or by activation through coupling to an environment.
%For small coupling to the environment, the activation increases with increasing coupling to the bath.
%For large coupling strengths, the activation rate is exponentially suppressed with the coupling strength.
%For the quantum trajectories, it seems that the coupling to the environment (which is the measurement strength $\gamma$ in our setup) first increases their ability to coherently explore longer distances.
%Only large measurement strengths suppress the coherence length.
%We observe a saturation for $\gamma \approx 2$, in agreement with the saturation of the coherence length of the mixed state (see main text and section 3).

\section{4. Liouvillian spectrum}

\begin{figure}[t!]
    \centering 
    \includegraphics[width=\columnwidth]{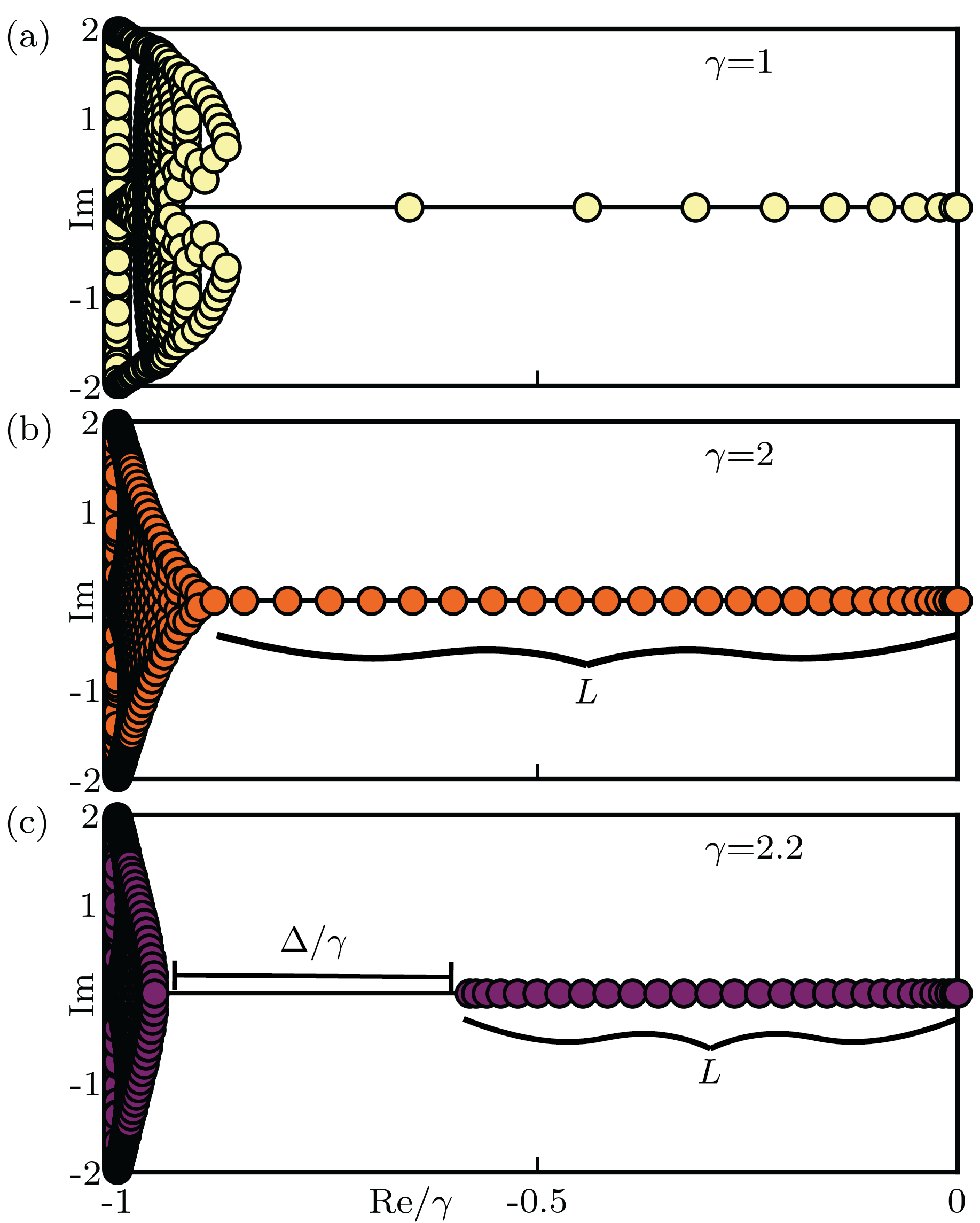}
    \caption{Complex Liouvillian spectrum of the single particle sector for $L=30$ and different values of the measurement strength $\gamma$.
    (a) With a finite measurement strength, the Liouvillian spectrum separates into a bulk of complex eigenvalues (left) and overdamped eigenvalues that collapse onto the real axis.
    There is a single steady state eigenvalue 0.
    With increasing measurement strength, more and more overdamped eigenvalues occur and ``evaporate'' onto the real axis.
    (b) At a critical measurement strength $\gamma=2$, the last overdamped eigenvalue collapses onto the real axis.
    There exist $L$ overdamped eigenvalues.
    (c) For measurement strength $\gamma \gtrsim 2$, a gap $\Delta$ opens between the complex eigenvalue bulk and the overdamped ones, marking a separation in lifetimes of the corresponding eigenmodes.
    }
    \label{fig: liouvillian spectrum}
\end{figure}

In the main text, we have introduced the Lindblad master equation governing the mixed state dynamics of our system,

\begin{equation}
    \label{eq: appendix Lindblad}
    \partial_t \rho = \mathcal{L}[\rho] =  -i [H, \rho] + \gamma \sum_{i=1}^L \left( n_i \rho n_i - \frac{1}{2}\{ n_i, \rho\}\right)\, ,
\end{equation}
with the Liouvillian superoperator $\mathcal{L}$ and a hopping Hamiltonian
\begin{equation}
\label{eq: appendix Hamiltonian}
    H~=\frac{J}{2}~\sum_{i=1}^{L-1} a_i^\dag a_{i+1} + a_{i+1}^\dag a_i\, .
\end{equation}
In terms of the eigenmodes $\sigma_i$ of the Liouvillian $\mathcal{L}$ with $\mathcal{L}[\sigma_i] = \xi_i \sigma_i$, the time evolution of the mixed state is given as
\begin{equation}
    \rho(t) = \sum_i e^{\xi_i t} {\rm Tr} (\sigma_i^\dag \rho ) \sigma_i \, .
\end{equation}
The eigenvalues $\xi_i$ are in general complex and have a negative real part ${\rm Re}\xi_i \leq 0$, which defines the lifetime of the eigenmode.
Conversely, the imaginary part ${\rm Im}\xi_i$ defines an oscillating phase. 

As discussed in the main text, the system~\eqref{eq: appendix Lindblad} has a unique steady-state with eigenvalue $\xi_0=0$.
The steady-state is the infinite temperature state, $\rho_0 \propto \mathds{1}$.
Any initial state $\rho(t=0)$ will evolve (up to symmetry restrictions) into the steady-state.
The eigenvalues $\xi_{i>0}$ describe how the initial state approaches the steady state.
The eigenvalue $\xi_1$ with the largest negative real value belongs to the longest living mode $\sigma_1$.
Indeed, the real part ${\rm Re}\xi_1$ is called the \emph{Liouvillian gap} and defines the maximal lifetime of any mode apart from the steady-state. 
We call modes that have eigenvalues with a vanishing imaginary part, ${\rm Im} \xi_i = 0$, \emph{overdamped} because their approach to the steady state is only governed by a lifetime, and no oscillations. 

The solution of the Lindblad equation~\eqref{eq: appendix Lindblad} can be found analytically, e.g., using a combination of Laplace and Fourier transforms~\cite{kenkre_brown_1985}.
More recently, the Lindblad equation was mapped to an imaginary-$U$ Hubbard model and solved by means of the Bethe Ansatz method~\cite{medvedyeva_et_al_2016}. 
However, from these exact methods it is challenging to gain intuition about the spectrum of the Liouvillian.
Instead, in Ref.~\cite{znidaric_2015}, the problem is restricted to the single-particle sector, alongside the Redfield approximation.
Thus, analytical expressions are found for the $L$ overdamped eigenvalues.

Here, we suffice to use a numerical evaluation that reveals a bulk of $\mathcal{O}(L^2)$ eigenvalues with ${\rm Re} \xi_i \gtrsim -\gamma$, and a maximum of $L$ overdamped eigenvalues with ${\rm Im} \xi_i = 0$ and $0\geq {\rm Re}\xi_i \geq -\gamma$~\cite{znidaric_2015}, see the Liouvillian spectrum of the single particle sector in Fig.~\ref{fig: liouvillian spectrum}.
To understand the formation of bulk eigenvalues and overdamped eigenvalues, it is illustrative to consider the two extreme cases $\gamma=0$ and $\gamma\to \infty$.
For $\gamma=0$, the Liouvillian~\eqref{eq: appendix Lindblad} has $L^2$  eigenvalues $\xi_{ij} = i(\omega_i - \omega_j)$, $i,j=1,\dots,L$, with $\omega_i$ the eigenvalues of the Hamiltonian~\eqref{eq: appendix Hamiltonian}.
Those eigenvalues have an infinite lifetime (${\rm Re}\xi_{ij}=0$) because there is no damping by $\gamma$.
Conversely, for $\gamma \to \infty$, the von Neumann term of the Lindblad equation~\eqref{eq: appendix Lindblad} vanishes, and we find two types of eigenmodes: (i) $L$ eigenmodes $\sigma_i=|i\rangle \langle i|$, $i=1,\dots, L$ and $|i\rangle = a_i^\dag |0\rangle$, with vanishing eigenvalues $\xi_i=0$; (ii) a bulk of $L(L-1)$ eigenmodes $\sigma_{ij}=|i\rangle \langle j|$, $i\neq j=1,\dots, L$, with eigenvalues $\xi_i=-\gamma$.
Between the two limits, there has to be a transition that separates the bulk eigenvalues from the eigenvalues with real part 0.
Specifically, at $\gamma\approx 2$ the last of the overdamped eigenvalues ``evaporates'' onto the real line.
By further increase of $\gamma$, a gap $\Delta$ opens up between the bulk of the eigenvalues and the overdamped ones.
This gap opening at $\gamma\approx 2$ is in agreement with our finding that the coherence length $l_\phi$ saturates around this value of the measurement strength.
In the thermodynamic limit $L \rightarrow \infty$, the overdamped eigenvalues become dense around $0$, meaning that the Liouvillian gap tends to zero.
In that case, one observes an algebraic rather than an exponential approach to the steady-state~\cite{cai_barthel_2013, znidaric_2015}.

 \section{5. Underdamped-to-overdamped transition for a single particle on two sites}
\begin{figure}[t!]
    \centering 
    \includegraphics[width=\columnwidth]{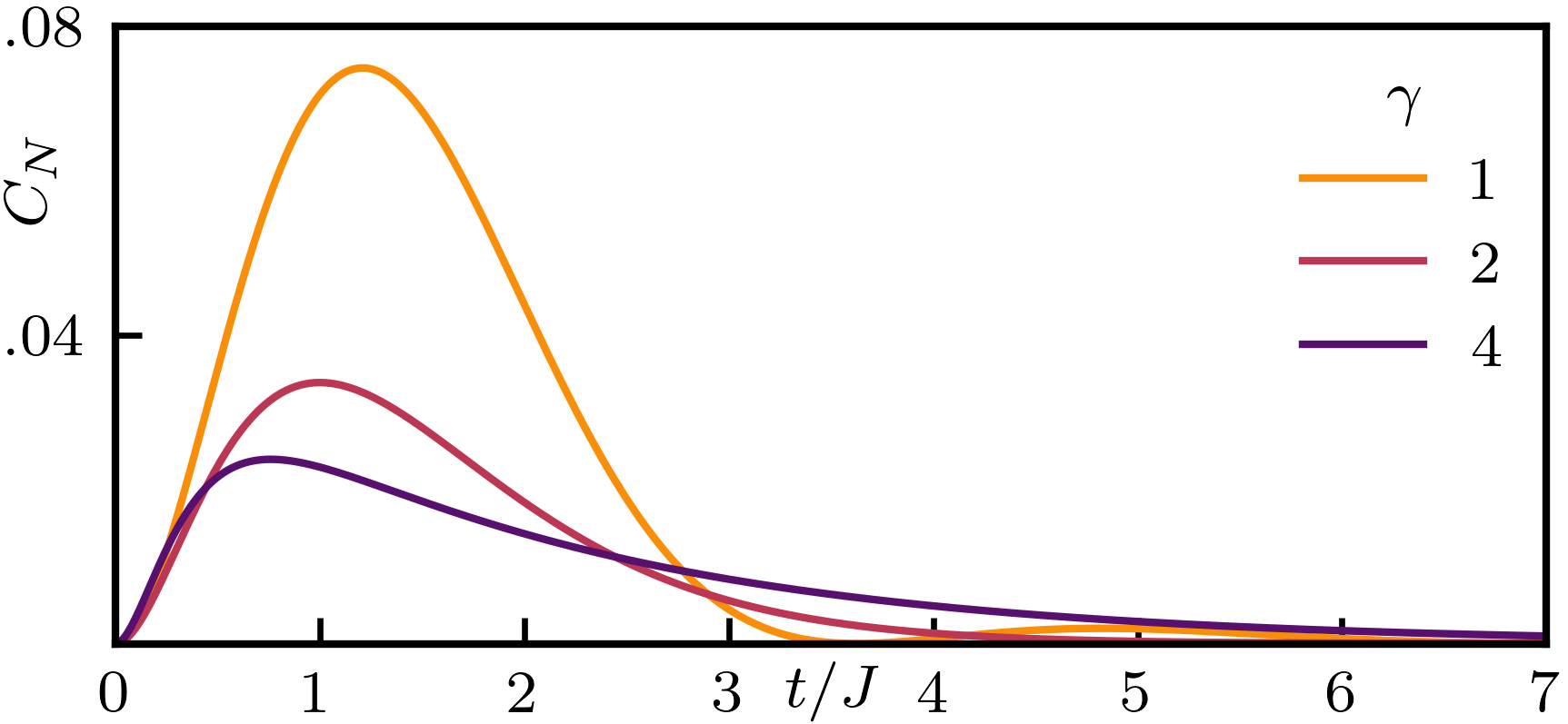}
    \caption{Configuration coherence for a single particle on two sites initially localized on site 1.
    The configuration coherence $C_N$ is initially zero in agreement with the separable initial state.
    At short times, it grows quadratically and then approaches the steady-state value of zero in 3 different ways: underdamped for $\gamma<2$, critically damped for $\gamma=2$, and overdamped for $\gamma>2$.
    }
    \label{fig: single particle transition}
\end{figure}

 In this section, we derive the underdamped-to-overdamped transition for a single particle on two sites subject to the Lindblad dynamics~\eqref{eq: appendix Lindblad}.
 The transition happens for a critical value of the measurement strength, $\gamma=2$ and is visible in the time evolution of the configuration coherence.
 We restrict the density matrix to the single particle sector,
 \begin{equation}
     \rho = \begin{bmatrix}
\alpha & \xi \\
\xi^* & \beta
\end{bmatrix},
 \end{equation}
where the probabilities $\alpha$ ($\beta$) of the particle being in the first (second) site have conditions $\alpha, \ \beta \in \mathbb{R}_{\geq 0}$ and $\tr\rho = \alpha + \beta = 1$.
The configuration coherence between the first and the second site is $\mathcal{C}_N=2|\xi|^2$.
Initially, we inject the particle into the first site, 
\begin{equation}
     \rho(t=0) = \begin{bmatrix}
1 & 0 \\
0 & 0
\end{bmatrix}.
 \end{equation}
For $J=1$, the Hamiltonian~\eqref{eq: appendix Hamiltonian} is given by
 \begin{equation}
     H = \begin{bmatrix}
0 & \frac{1}{2} \\
\frac{1}{2} & 0
\end{bmatrix}.
 \end{equation}
 The Lindblad equation in matrix form becomes
\begin{equation}
    \partial_t \begin{bmatrix}
\alpha & \xi \\
\xi^* & \beta
\end{bmatrix}
= \begin{bmatrix}
i(\xi/2-\xi^*/2) & i(\alpha/2 - \beta/2)-\gamma \xi \\
i( \beta/2-\alpha/2) -\gamma \xi^* &i(\xi^*/2-\xi/2)
\end{bmatrix}.
\end{equation}
Note that these are exactly the rate equations of aligned double quantum dots measured by a point contact detector~\cite{gurvitz_1997}.
We are interested in the off-diagonal element $\xi$, and performing a second time derivative, we decouple the differential equations and obtain
\begin{equation}
    \partial_t^2 \xi = \frac{\xi^*-\xi}{2} - \gamma \partial_t \xi \, .
\end{equation}
The real part of $\xi$ is therefore an exponential, and the imaginary part describes a damped harmonic oscillator (which, of course, is also exponentially decaying).
With our initial conditions $\xi(t=0)=0$ and $\partial_t \xi (t=0) = i(\alpha(t=0)/2 - \beta(t=0)/2) - \gamma \xi(t=0) = i/2$, we find ${\rm Re}\xi(t)=0$ and that the imaginary part undergoes an underdamped-to-overdamped transition for $\gamma=2$.
For the configuration coherence, this yields
\begin{widetext}
\begin{equation}
    C_N(t) = 2|{\rm Im}\xi(t)|^2 = 
    \begin{cases}
        \frac{e^{-\gamma t}}{2\,(1-(\gamma/2)^2)}\sin^2\left(\sqrt{1-(\gamma/2)^2}t\right), \ \gamma<2 \ {\rm (underdamped)} \\
        \frac{1}{2}t^2e^{-2t}, \ \gamma = 2 \ {\rm (critical \ damping)} \\
        \frac{e^{-\gamma t}}{2\,((\gamma/2)^2 - 1)}\sinh^2\left(\sqrt{(\gamma/2)^2 - 1}t\right), \ \gamma > 2 \ {\rm (overdamped)} .
        
    \end{cases}
\end{equation}
\end{widetext}
Fig.~\ref{fig: single particle transition} shows how the configuration coherence approaches its steady-state value 0 in the 3 regimes.
For short times $t\ll 1/\gamma$, all regimes show quadratic increase, $C_N(t) \approx t^2/2$.

\section{6. Configuration coherence for fermionic Gaussian states}
In this section, we explain how to obtain the configuration coherence for mixtures of fermionic Gaussian states.
We start from an individual quantum trajectories' Gaussian state matrix that can be efficiently simulated along the stochastic evolution~\cite{cao_et_al_2019}.
The Gaussian state of $N$ fermionic particles on $L$ sites is described by means of the state matrix $U_{jk}$,
\begin{equation}
    \state{\psi} = \prod_{j=1}^N \left( \sum_{k=1}^L U_{jk} a_j^\dag \right) \state{0} \, .
\end{equation}

As an illustrative example, we consider $N=2$ particles on $L=4$ sites.
Using the fermionic commutation relations $\{a_j, a_k^\dag\}=\delta_{jk}$, we obtain the following state from the entries of the matrix $U$,
\begin{align}
\label{eq: appendix Gaussian state}
    \state{\psi} = &(U_{13}U_{24} - U_{14}U_{23})\state{0011}\\
     + &(U_{12}U_{24} - U_{14}U_{22})\state{0101}\nonumber\\
     + &(U_{12}U_{23} - U_{13}U_{22})\state{0110}\nonumber\\
     + &(U_{11}U_{24} - U_{14}U_{21})\state{1001}\nonumber\\
     + &(U_{11}U_{23} - U_{13}U_{21})\state{1010}\nonumber\\
     + &(U_{11}U_{22} - U_{12}U_{21})\state{1100} \, .\nonumber
\end{align}
Note that the amplitudes are determinants of all possible $2\times 2$ matrices formed by choosing two columns of the $2\times 4$ matrix $U$.
This generalizes to arbitrary numbers of particles $N$ and chain lengths $L$, where the prefactors are all possible determinants of $N\times N$ matrices formed by choosing $N$ columns of the $N\times L$ matrix $U$.

For a bipartition of the chain into two subsystems, the choice of columns reflects the configuration of the $N$ particles w.r.t. the cut.
As an example we consider the prefactor of $\state{0101}$ in the state~\eqref{eq: appendix Gaussian state} and a bipartition of the chain in the middle, i.e., sites 1 and 2 form subsystem $A$, and sites 3 and 4 form subsystem $B$.
The corresponding determinant $U_{12}U_{24} - U_{14}U_{22}$ is the determinant of the $2\times 2$ matrix
\begin{equation}
\begin{bmatrix}
    U_{12} & U_{14} \\
    U_{22} & U_{24}
\end{bmatrix} \, ,
\end{equation}
which we obtain by choosing the first column (corresponding to site 2 in subystem $A$) and the last column (corresponding to site 4 in subsystem $B$) of the matrix $U$.
Let $L_A$ ($L_B$) be the size of subsystem $A$ ($B$).
For a configuration of $0\leq n\leq N$ particles in subsystem $A$, there are ${L_A \choose n} \times {L_B \choose N-n}$ many determinants, which we label $|U|_{n,i}$, $i=1,\dots,{L_A \choose n} \times {L_B \choose N-n}$.

The $N$-particle Fock block of the pure state's density matrix $\rho = \pure{\psi}$ is an ${L \choose N} \times {L \choose N}$ matrix with entries being products of determinants and complex conjugate determinants.
To obtain the mixed state $\bar{\rho}=1/M\sum_{i=1}^M \pure{\psi_i}$ of Gaussian quantum trajectories, one averages over such products of determinants.
The configuration coherence is then calculated as
\begin{equation}
\label{eq: appendix configuration coherence}
    C_N(\bar{\rho}) = 2\sum_{m>n=0}^N \sum_{i, j} \overline{|U|_{n,i} |U|_{m,j}^*}\, .
\end{equation}
This expression allows to calculate the mixed state entanglement of fermionic many-body states of an extensive number of particles on large systems.

\end{document}